\documentclass[aps,prd,twocolumn,showpacs,nofootinbib]{revtex4}

\usepackage{graphicx}
\usepackage{dcolumn}

\begin{document}


\title{QCD based quark description of $\pi$-$\pi$ scattering
up to the $\sigma$ and $\rho$ region}


\author{Stephen R. Cotanch}
\email[]{cotanch@ncsu.edu}

\author{Pieter Maris}
\email[]{pmaris@unity.ncsu.edu}
\affiliation{Department of Physics,
North Carolina State University, Raleigh,  NC 27695-8202}

\date{\today}

\begin{abstract}
We study forward and backward $\pi$-$\pi$ scattering within a QCD
model based on the Dyson--Schwinger, Bethe--Salpeter equations
truncated to the rainbow-ladder level.  Our microscopic relativistic
quark formulation preserves chiral symmetry and reproduces the
observed scattering lengths for total isospin zero, one and two.  
At higher energies both scalar and vector meson resonances naturally
occur in the scattering amplitudes.  We also report a comparative
study with phenomenological meson-exchange models and find such
approaches are reasonable especially near $\pi$-$\pi$ resonances.
\end{abstract}



\maketitle

\section{\label{sec:intro}
Introduction}

As the lightest hadron, the Goldstone boson of spontaneous chiral
symmetry breaking and dominant particle governing the nucleon-nucleon
interaction, the pion occupies a special position in hadronic physics.
Accordingly, $\pi$-$\pi$ scattering has attracted considerable
interest even though the cross section is not directly measurable.  In
particular, a variety of theories have been utilized to make
scattering predictions: at low energies, current
algebra/PCAC~\cite{Weinberg:1966kf}, chiral perturbation
theory~\cite{Weinberg:1967fm,Weinberg:1979kz,Gasser:1984yg,Donoghue:1988xa}
and bosonization models with pion fields~\cite{Roberts:1994ks};
Quantum Hadrodynamic [QHD] meson exchange models and unitarized
relativistic coupled channels~\cite{Pichowsky:2001qe} for intermediate
energies; Regge theory at moderately higher energies; perturbative QCD
at very high energies.  These in turn have been confronted by
experimental phase shift
analyses~\cite{Grayer:1974cr,Srinivasan:1975tj,Rosselet:1977pu,Kaminski:1999ea}
yielding reasonable agreement.  However, there have been very few
non-perturbative, relativistic covariant quark predictions which is
the thrust of this paper.

This work applies the quark Dyson--Schwinger [DSE] and meson
Bethe--Salpeter [BSE]
equations~\cite{Roberts:1994dr,Roberts:2000aa,Alkofer:2000wg}, in the
rainbow-ladder approximation, to $\pi$-$\pi$ scattering and extends
our previous threshold analysis~\cite{Bicudo:2001jq} which reproduced
Weinberg's low energy theorem~\cite{Weinberg:1966kf} and the Adler
zero.  The approach is QCD based, renormalizable, relativistic,
rigorously covariant, embodies crossing and chiral symmetry, and
contains only two predetermined parameters that provide a realistic,
comprehensive description of the light meson spectra and
decays~\cite{Maris:1997tm,Maris:1999nt,Maris:1999bh,Maris:2000sk,%
Jarecke:2002xd}.  Our key finding is that the
model correctly predicts all three isospin amplitudes measured at low
energies and also reproduces the observed $\sigma$ and $\rho$
resonances at higher energies. Our framework can thus assess
phenomenological meson exchange models which require resonances as
input and we find such models are quite reasonable.

This paper is organized into four sections.
Section~\ref{sec:DSEapproach} formulates the scattering problem and
addresses the main approximation: the rainbow truncation of the DSE
for the quark propagator and the ladder truncation of the BSE for the
pion as a quark-antiquark bound state.  This section also details the
quark annihilation and exchange diagrams and relates the amplitudes
for these processes to the various isospin scattering amplitudes for
total $\pi$-$\pi$ isospin zero, one and two.  Section~\ref{sec:numres}
presents large scale, supercomputer calculations for forward and
backward scattering and compares results to an effective meson
exchange model where the coupling constants and pole masses are a
prerequisite. Finally, conclusions are summarized in section IV and
supporting mathematical details are provided in an appendix.

\section{\label{sec:DSEapproach}
Meson scattering in the Dyson--Schwinger approach}

\subsection{\label{subsec:DSEs}
Dyson--Schwinger Equations}
The DSE for the renormalized quark propagator having four momenta $p$
in Euclidean space is
\begin{eqnarray}
S(p)^{-1} &=& i \, Z_2\, /\!\!\!p + Z_4\,m_q(\mu) +
\nonumber \\ && {}
        Z_1 \int_q^\Lambda\! g^2 D_{\mu\nu}(k)
	        \textstyle{\frac{\lambda^{\alpha}}{2}}
 		\gamma_\mu S(q)\Gamma^{\alpha}_\nu(q,p) \,,
\label{gendse}
\end{eqnarray}
where $D_{\mu\nu}(k)$ is the dressed-gluon propagator,
$\Gamma^{\alpha}_\nu(q,p)$ the dressed-quark-gluon vertex with color
index ${\alpha} = 1...8$, and \mbox{$k=p-q$}. The notation
$\int_q^\Lambda\!$ represents $\int_q^\Lambda\!\frac{d^4k}{(2\pi)^4}$.
The most general propagator solution of Eq.~(\ref{gendse}) has the
form \mbox{$S(p)^{-1} = i /\!\!\! p A(p^2) +$} \mbox{$B(p^2)$} and is
renormalized at spacelike $\mu^2$ according to \mbox{$A(\mu^2)=1$} and
\mbox{$B(\mu^2)=m_q(\mu)$} with $m_q(\mu)$ being the current quark mass.  
We use the Euclidean metric where \mbox{$\{\gamma_\mu,\gamma_\nu\} =
2\delta_{\mu\nu}$}, \mbox{$\gamma_\mu^\dagger = \gamma_\mu$} and
\mbox{$a\cdot b = a_i b_i \equiv \sum_{i=1}^4 a_i b_i$}.

Mesons are described by the Bethe--Salpeter amplitude [BSA],
$\Gamma_H$, which is a solution of the homogeneous BSE for $q^a
\bar{q}^b$ bound states given by
\begin{eqnarray}
 \Gamma^{a\bar{b}}_H(p_+,p_-) &=&
        \int_q^\Lambda\! K(p,q;P)
\nonumber \\ && {}
        \times S^a(q_+) \, \Gamma^{a\bar{b}}_H(q_+,q_-) \, S^b(q_-)\, ,
\label{homBSE}
\end{eqnarray}
where $a$ and $b$ are flavor indices, $p_+ = p + \eta P$ and $p_- = p
- (1-\eta) P$ are the outgoing and incoming quark momenta
respectively, and $q_\pm$ is defined similarly.  The kernel $K$ is the
renormalized, amputated $q\bar q$ scattering kernel that is
irreducible with respect to a pair of $q\bar q$ lines.  This equation
has solutions at discrete values of $P^2 = -m_H^2$, where $m_H$ is the
meson mass.  By then imposing the canonical normalization condition
for $q\bar q$ bound states, $\Gamma_H$ is uniquely determined.  Mesons
with different spin, parity and C parity, such as pseudo-scalar,
vector, etc., are characterized by different Dirac structures.  The
most general decomposition for pseudoscalar bound states
is~\cite{Maris:1997tm}
\begin{eqnarray}
\label{genpion}
\lefteqn{\Gamma_{PS}(q_+,q_-) = \gamma_5 \big[ i E(q^2,q\cdot P;\eta)
        + \;/\!\!\!\! P \, F(q^2,q\cdot P;\eta) }
\nonumber \\ && {}
        + \,/\!\!\!k \, G(q^2,q\cdot P;\eta)
        + \sigma_{\mu\nu}\,P_\mu q_\nu \,H(q^2,q\cdot P;\eta) \big]\,,
\end{eqnarray}
where the invariant amplitudes $E$, $F$, $G$ and $H$ are Lorentz
scalar functions of $q^2$ and $q\cdot P$.  For eigenstates of C
parity, these amplitudes are appropriately odd or even in the C odd
variable $q\cdot P$.  In the case of the $0^{-+}$ pion, for example,
the amplitude $G$ is odd in $q\cdot P$, the others are even.  Note
also that these amplitudes explicitly depend on the momentum
partitioning parameter $\eta$.  However, provided Poincar\'e
invariance is preserved, the resulting physical observables are $\eta$
independent~\cite{Maris:1997tm,Maris:1999nt,Maris:2000sk}.

\subsection{\label{subsec:rainbowladder}
Rainbow-ladder truncation}
We use the rainbow-ladder truncation for the system of
Dyson--Schwinger and Bethe--Salpeter equations [DSBS].  In particular,
the rainbow truncation of the quark DSE, Eq.~(\ref{gendse}), is
\begin{equation}
\label{ourDSEansatz}
Z_1 \, g^2 D_{\mu \nu}(k) \Gamma^{\alpha}_\nu(q,p) \rightarrow
 	{\cal G}(k^2) D_{\mu\nu}^{\rm free}(k)\, \gamma_\nu
	\textstyle{\frac{\lambda^{\alpha}}{2}} \,,
\end{equation}
where $D_{\mu\nu}^{\rm free}(k=p-q)$ is the free gluon propagator in
Landau gauge, and ${\cal G}(k^2)$ is an effective $\bar q q$
interaction that reduces to the perturbative QCD running coupling in
the ultraviolet region.  The corresponding ladder truncation of the
BSE, Eq.~(\ref{homBSE}), is
\begin{equation}
\label{ourBSEansatz}
        K(p,q;P) \to
        -{\cal G}(k^2)\, D_{\mu\nu}^{\rm free}(k)
        \textstyle{\frac{\lambda^{\alpha}}{2}}\gamma_\mu
        \textstyle{\frac{\lambda^{\alpha}}{2}}\gamma_\nu \,,
\end{equation}
where \mbox{$k=p-q$}.  The two truncations combine to consistently
produce vector and axial-vector vertices satisfying the respective
Ward--Takahashi identities.  In the axial case, this ensures that in
the chiral limit the ground state pseudoscalar mesons are the massless
Goldstone bosons associated with chiral symmetry
breaking~\cite{Maris:1997tm,Maris:1998hd}.  For vector mesons it
yields a conserved electromagnetic current if the impulse
approximation is used to calculate the electromagnetic form
factor~\cite{Maris:2000sk}.  Furthermore, this truncation was found to
be particularly suitable for the flavor octet pseudoscalar and vector
mesons since the next-order corrections in a quark-gluon skeleton
graph expansion significantly cancel~\cite{Bender:1996bb}.

\subsection{\label{subsec:pipi}
$\pi$-$\pi$ scattering}
There are different types of diagrams that contribute to $\pi$-$\pi$
scattering, or more general, meson-meson scattering, such as quark
annihilation or quark exchange diagrams.  It is convenient to first
discuss these different types of contributions in the impulse
approximation, even though the impulse approximation is known to be
insufficient~\cite{Bicudo:2001jq}.  The more correct treatment is
given in the next subsection.

\begin{figure}[h]
\includegraphics[width=8cm]{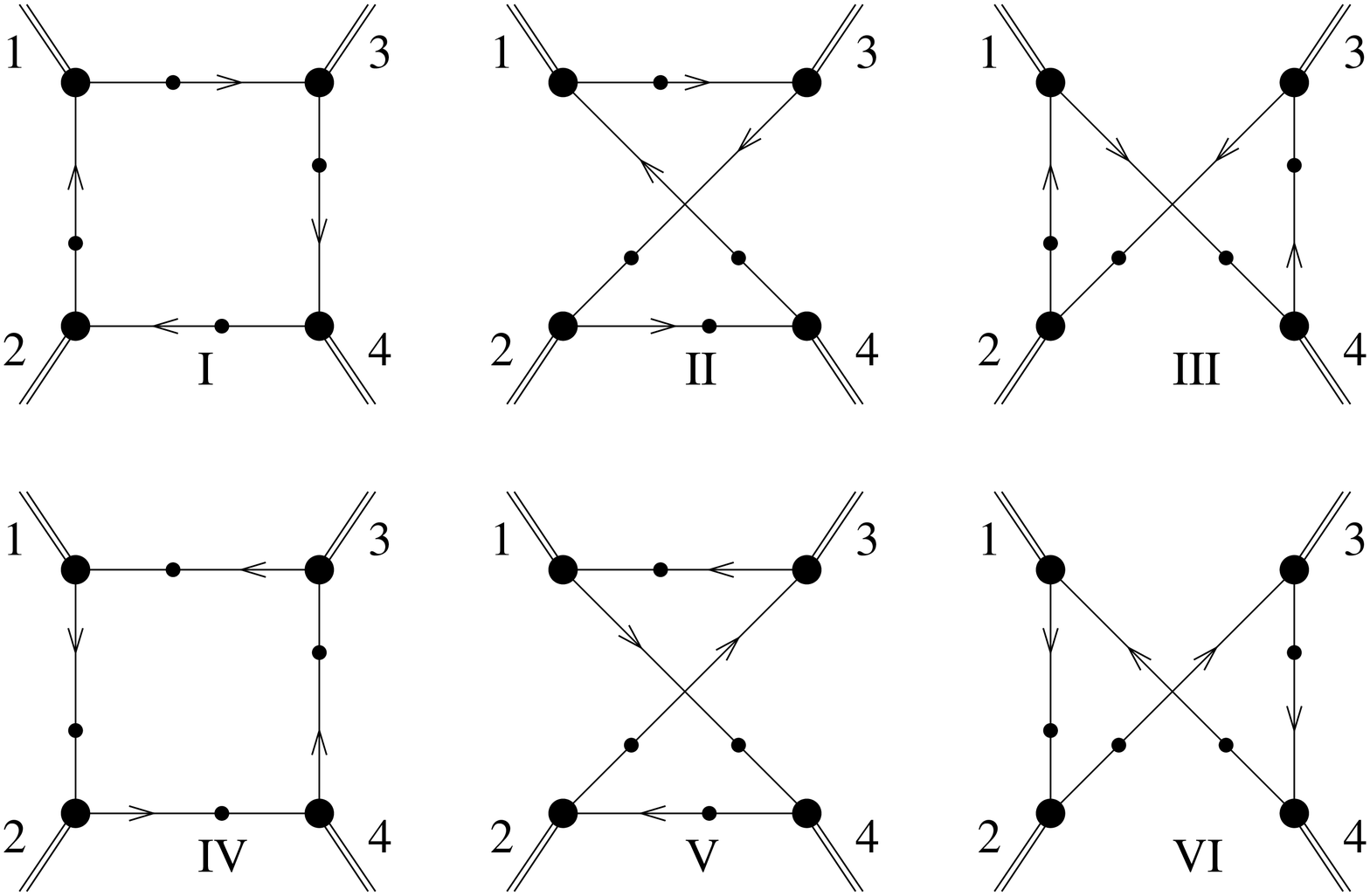}
\caption{\label{fig:conf}
The six diagrams contributing to the impulse amplitude for $\pi$-$\pi$
scattering; the incoming pions are labeled 1 and 2, and the outgoing
pions are 3 and 4.}
\end{figure}

Figure~\ref{fig:conf} depicts the entire class of diagrams that
constitute the impulse approximation for the scattering amplitude.
Each diagram can be associated with a specific amplitude that is a
function of the Mandelstam variables $s = (P_1 + P_2)^2$, $t = (P_1 -
P_3)^2$, and $u = (P_1 - P_4)^2$, where $P_1$, $P_2$ ($P_3$, $P_4$)
are the incoming (outgoing) pion momenta.  The external pions are
on-shell, yielding $P_i^2 = -m_\pi^2$ in the Euclidean metric.  At
threshold $s= -4m_\pi^2$, $t=u=0$, and the physical region has $s <
-4m_\pi^2$.  Momentum conservation requires $s + t + u = -4 m_\pi^2$,
so that the amplitudes, which we chose to express as functions of $s$,
$t$, and $u$, only depend on two independent variables.  Alternatively
the amplitudes could be represented as function of the pion
center-of-momentum [c.m.] energy $\omega = \sqrt{-s}/2$ and the c.m.
scattering angle $\theta$ with $\cos(\theta) = (u - t)/(u + t)$.  Then
forward (backward) scattering, $\theta=0$ ($\theta=180$), corresponds
to $t=0$ ($u=0$).

Because there are only two distinct microscopic scattering mechanisms,
quark annihilation and quark exchange, there are actually only two
distinct amplitudes, $D$ and $E$, respectively.  Thus, all six
diagrams in Fig.~\ref{fig:conf} can be represented by these two
\begin{eqnarray}
	{\hbox {diagram I:}} &\;\;& 	D(s,t,u)	\\
	{\hbox {diagram II:}} &\;\;& 	E(s,t,u)	\\
	{\hbox {diagram III:}} &\;\;& 	D(s,u,t)	\\
	{\hbox {diagram IV:}} &\;\;& 	D(s,t,u)	\\
	{\hbox {diagram V:}} &\;\;& 	E(s,u,t)	\\
	{\hbox {diagram VI:}} &\;\;& 	D(s,u,t) \;.
\end{eqnarray}
These amplitudes can be calculated from integrals involving the quark
propagators and pion BSAs, e.g.
\begin{eqnarray}
\lefteqn{D(s,t,u) =
	2\;N_c \int_q^\Lambda\! {\rm Tr}\Big[ 
	S(k+P_1) \Gamma_\pi(k+P_1,k) S(k) }
\nonumber \\ && {} \times
	\Gamma_\pi(k,k-P_2) S(k-P_2) 
	\bar{\Gamma}_\pi(k-P_2,k+P_1-P_3) 
\nonumber \\ && {} \times
	S(k+P_1-P_3) \bar{\Gamma}_\pi(k+P_1-P_3,k+P_1)]	\,.
\end{eqnarray}
Note that for $\pi$-$\pi$ scattering, $E$ is symmetric under
interchange of $u$ and $t$ (indicated hereafter by a semicolon),
$E(s;u,t) = E(s;t,u)$, whereas $D$ is symmetric in $s$ and $t$,
$D(s,t;u) = D(t,s;u)$.  Furthermore, one can relate $E$ to $D$:
$E(s;t,u) = D(u,t;s)$.  However, we find it more convenient to retain
and calculate both $D$ and $E$ independently since this affords a
sensitive numerical check on our computer codes.

The three isospin amplitudes, $T_I$, for $\pi$-$\pi$ scattering are
specific combinations of these amplitudes (see the appendix for
further details).
\begin{eqnarray}
	T_0(s,t,u) &=& 3 (D(s,t;u) + D(s,u;t))	
\nonumber \\ && {}
	- \frac{1}{2}(E(s;t,u) + E(s;u,t))	\;,
\label{eq:isospinT0}	\\
	T_1(s,t,u) &=& 2 (D(s,t;u) - D(s,u;t))	\;,
\label{eq:isospinT1}	\\
	T_2(s,t,u) &=& E(s;u,t) + E(s;t,u)	\;.
\label{eq:isospinT2}
\end{eqnarray}
Crossing symmetry and Bose statistics requires the even isospin
amplitudes to be symmetric in exchange of $u$ and $t$ and only contain
even partial waves.  Similarly the isospin one amplitude must be
antisymmetric in interchange of $u$ and $t$ and have only odd partial
waves.  Hence at threshold $T_1 (-4 m^2_{\pi},0,0) = 0$ but the
isospin zero and two amplitudes $T_0$ and $T_2$ are nonzero.

\subsection{\label{subsec:beyond}
Beyond the impulse approximation}
\begin{figure}[b]
\includegraphics[width=8.5cm]{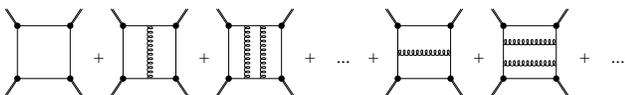}
\caption{\label{fig:beyond}
Diagrams needed to correctly describe $\pi$-$\pi$ scattering
in the DSBS approach in rainbow-ladder truncation.}
\end{figure}

As documented in Ref.~\cite{Bicudo:2001jq}, the impulse approximation
combined with the rainbow-ladder truncation is insufficient to
describe $\pi$-$\pi$ scattering and it is necessary to include the
ladder kernel as indicated in Fig.~\ref{fig:beyond}.
Thus, for amplitude $D$ we need to calculate and add an infinite
series of $s$-channel ladder diagrams as well as an infinite series of
$t$-channel ladder diagrams to the impulse term. Similarly for
amplitude $E$, we must add and compute an infinite series of
$t$-channel and $u$-channel ladder diagrams.

These infinite set of ladder diagrams can be calculated by solving an
inhomogeneous BSE in ladder truncation
\begin{eqnarray}
\lefteqn{ F(p,P_i,P_j) =  F_0(p,P_i,P_j) +
	\int^\Lambda_q \!{\cal G}(k^2)\, D_{\mu\nu}^{\rm free}(k) }
\nonumber \\ && {}
        \times 
	\textstyle{\frac{\lambda^{\alpha}}{2}} \gamma_\mu 
	S(q_+) \, F(q,P_i,P_j)\, S(q_-) 
        \textstyle{\frac{\lambda^{\alpha}}{2}} \gamma_\nu
\label{eq:inhomBSE}
\end{eqnarray}
where now $q_\pm = q \pm (P_i + P_j)$ for $i,j=1,2$ ($s$-channel
ladders) or $q_\pm = q \pm (P_i - P_j)$ for $i=1,2$ and $j=3,4$ ($t$-
and $u$-channel ladders).  Here, $F_0(p,P_i,P_j)$ is an inhomogeneous
term of the type $\Gamma_\pi \, S \, \Gamma_\pi$; e.g. for the
$s$-channel ladder contribution to configuration $D$ we have $i=1$,
$j=2$ and
\begin{eqnarray}
F_0(p,P_1,P_2) &=&
	\Gamma_\pi(p+Q_+,p-Q_-) \, S(p-Q_-)
\nonumber \\ &&	{} \times
	\Gamma_\pi(p-Q_-,p-Q_+)\,  
\end{eqnarray}
where $Q_\pm = (P_1 \pm P_2)/2$.  This is similar to solving the
inhomogeneous BSE for the quark-photon vertex, as done in
Refs.~\cite{Maris:1999bh,Maris:2000sk}.  The Dirac structure is indeed
of equal complexity, in both cases there are eight independent terms.
Thus, $F$ can be written as
\begin{eqnarray}
	F(p,P_i,P_j) &=& \sum_n O_n \;
			f_n(p^2,p\cdot P_i,p\cdot P_j)
\end{eqnarray}
with
\begin{eqnarray}
	O_1(p,P_i,P_j) &=& 1			\\
	O_2(p,P_i,P_j) &=& \gamma\cdot p	\\
	O_3(p,P_i,P_j) &=& \gamma\cdot P_i	\\
	O_4(p,P_i,P_j) &=& \gamma\cdot P_j	\\
	O_5(p,P_i,P_j) &=& 
		\gamma\cdot p \; \gamma\cdot P_i	\\
	O_6(p,P_i,P_j) &=& 
		\gamma\cdot p \; \gamma\cdot P_j	\\
	O_7(p,P_i,P_j) &=& 
		\gamma\cdot P_i \; \gamma\cdot P_j	\\
	O_8(p,P_i,P_j) &=& 
		\gamma\cdot p \; \gamma\cdot P_i \; \gamma\cdot P_j \;.
\end{eqnarray}
Note that the $f_n$ are functions of {\em three} independent
variables: $p^2$, $p\cdot P_i$ and $p\cdot P_j$, in contrast to the
quark-photon vertex, which has only two independent variables.

\section{\label{sec:numres}
Numerical results}

We consider only forward and backward scattering which simplifies the
numerical analysis compared to scattering for arbitrary angles.
However, since $F$ is a function of {\em three} independent variables
for any given $P_1$ and $P_2$, there is a significant increase in the
numerical effort to solve Eq.~(\ref{eq:inhomBSE}) when compared to the
inhomogeneous BSE for the quark-photon vertex.  We discretize the
three independent variables on a three-dimensional grid and solve by
iteration, starting with $F_0$.  If there are no singularities, the
number of iterations for convergence is about 20 to 30. The actual
calculations are done on a parallel processor supercomputer (IBM SP)
and scale reasonably well, typically 2 to 3 CPU minutes on 32
processors per external $P_i,P_j$.  Since there are two sets of ladder
diagrams for both configurations $D$ and $E$, about 5 CPU hours per
external momentum variable is required.  The iterative procedure is
significantly less efficient near a singularity, yet the entire
parallel calculation still scales reasonable well.

\subsection{\label{subsec:model}
Model truncation}
We adopt the model framework and parameters that have been recently
developed~\cite{Maris:1999nt} which provide a good description of the
masses and decay constants of the light pseudoscalar and vector
mesons.  The Ansatz for the effective $\bar q q$ interaction is
\begin{eqnarray}
\label{gvk2}
\frac{{\cal G}(k^2)}{k^2} &=&
        \frac{4\pi^2\, D \,k^2}{\omega^6} \, {\rm e}^{-k^2/\omega^2}
\nonumber \\ && {}
        + \frac{ 4\pi^2\, \gamma_m \; {\cal F}(k^2)}
        {\textstyle{\frac{1}{2}} \ln\left[\tau +
        \left(1 + k^2/\Lambda_{\rm QCD}^2\right)^2\right]} \;,
\end{eqnarray}
where \mbox{$\gamma_m=12/(33-2N_f)$} and \mbox{${\cal F}(s)=(1 -
\exp\frac{-s}{4 m_t^2})/s$}.  The ultraviolet behavior is chosen to be
the QCD running coupling $\alpha(k^2)$; the ladder-rainbow truncation
then generates the correct perturbative QCD structure for the DSBS
system of equations.  The first term implements the infrared strength
in the region \mbox{$0 < k^2 < 1\,{\rm GeV}^2$} where the chiral
condensate is fit~\cite{Hawes:1998cw}.  We use \mbox{$m_t=0.5\,{\rm
GeV}$}, \mbox{$\tau={\rm e}^2-1$}, \mbox{$N_f=4$}, \mbox{$\Lambda_{\rm
QCD} = 0.234\,{\rm GeV}$}, and a renormalization scale
\mbox{$\mu=19\,{\rm GeV}$} which is in the perturbative
domain~\cite{Maris:1997tm,Maris:1999nt}.  The remaining parameters,
\mbox{$\omega = 0.4\,{\rm GeV}$}, \mbox{$D=0.93\,{\rm GeV}^2$} and the
$u/d$ degenerate quark mass, are determined by fitting the chiral
condensate, $m_\pi$ and $f_{\pi}$.  The predicted $\rho$ meson mass
and electroweak decay constant are also in good agreement with
observation~\cite{Maris:1999nt}, as can be seen from
Table~\ref{tab:masses}.  Further, without parameter readjustment, the
model agrees remarkably well with the most recent Jlab
data~\cite{Volmer:2000ek} for the pion charge form factor $F_\pi(Q^2)$
and the strong decay $\rho \to \pi\pi$~\cite{Jarecke:2002xd}.
\begin{table}[h]
\caption{\label{tab:masses}
Calculated and measured meson masses and decay
constant, adapted from Ref.~\protect\cite{Maris:1999nt}.}
\begin{ruledtabular}
\begin{tabular}{l|llr}
        & experiment~\protect\cite{Groom:2000in}& calculated	&\\
        & (estimates)	& ($^\dagger$ fitted) 		&\\ \hline
$m^{u=d}_{\mu=1 {\rm GeV}}$ &	5 - 10 MeV	& 5.5 MeV     	&\\
- $\langle \bar q q \rangle^0_{\mu}$
                & (0.236 GeV)$^3$ & (0.241$^\dagger$)$^3$ &\\
$m_\pi$         &  0.1385 GeV &   0.138$^\dagger$ GeV	&\\
$f_\pi$         &  0.924 GeV  &   0.925$^\dagger$ GeV	&\\
$m_\rho$        &  0.770 GeV  &   0.742 GeV	&\\
$f_\rho$        &  0.153 GeV  &   0.146 GeV	&\\
\end{tabular}
\end{ruledtabular}
\end{table}
%

\subsection{Results for isospin amplitudes near threshold}
\begin{figure}[t]
\includegraphics[width=8cm]{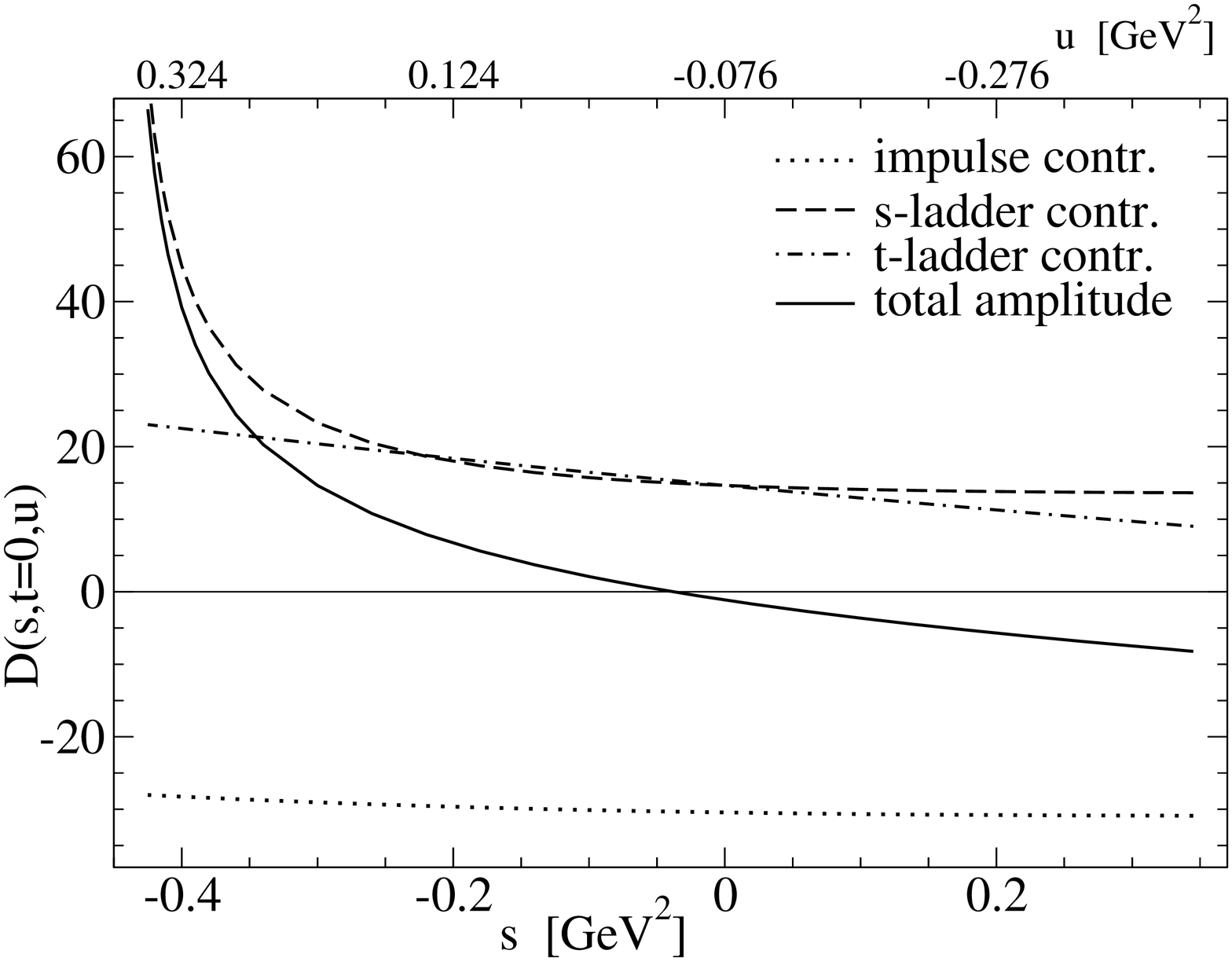}
\includegraphics[width=8cm]{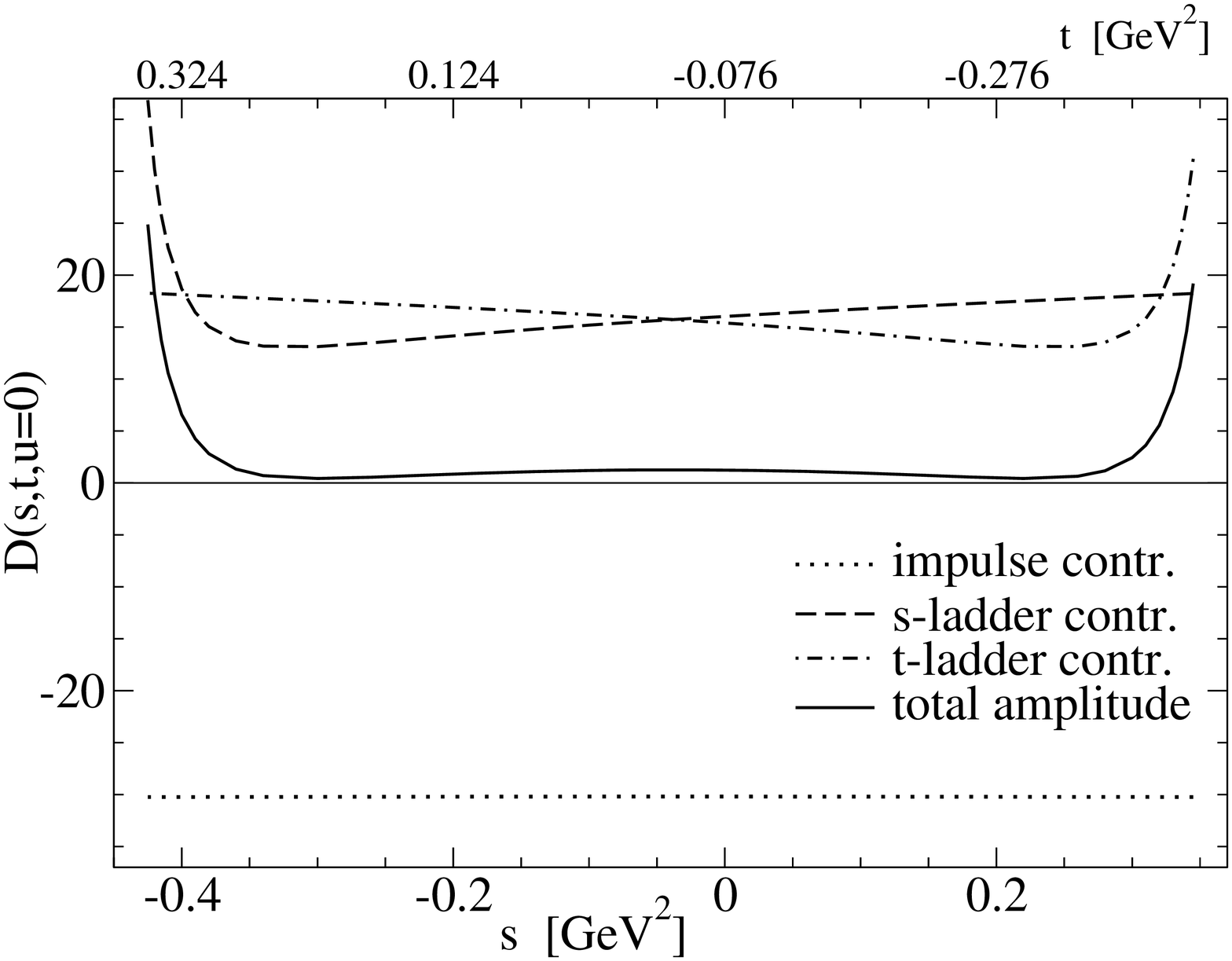}
\caption{\label{fig:ladder}
Numerical results for $D(s,t=0;u)$ (top) and for $D(s,t;u=0)$ (bottom),
using impulse approximation only (dotted), ladder contributions (dashed
and dot-dashed), and total (solid), which is the sum of the impulse
contribution and the two sets of ladder contributions.}
\end{figure}

We first elucidate the importance of the ladder contributions,
reaffirming that the impulse approximation is insufficient to describe
$\pi$-$\pi$ scattering.  In Fig.~\ref{fig:ladder} the impulse
approximation (dotted curve) is compared to the contributions from the
entire $s$-channel and $t$-channel ladder diagrams for amplitude
$D(s,t;u)$ at extreme forward and backward scattering angles.  The
total amplitude (solid curve) is noticeably different than the impulse
amplitude, clearly indicating the ladder exchange diagrams are
crucial.  Also note that $D(s,t;0)$ is indeed symmetric in
$s$ and $t$ as anticipated; within our numerical accuracy we also
found that $E(s;0,u)=E(s;u,0)=D(s,0;u)$.

\begin{figure}[t]
\includegraphics[width=8cm]{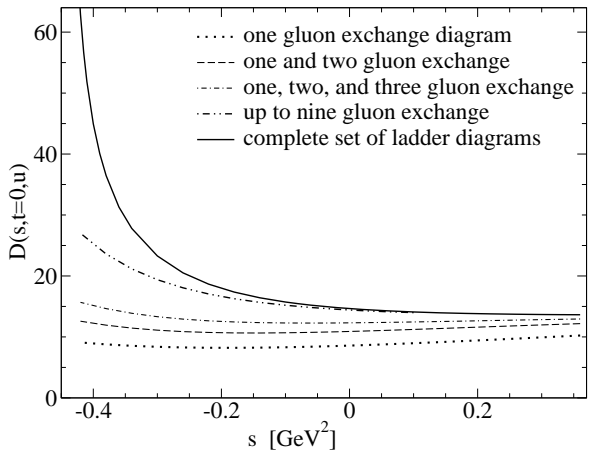}
\caption{\label{fig:129ladder}
Numerical results for contributions to $D(s,t=0;u)$
coming from the $s$-channel one gluon exchange diagram (dotted),
one and two gluon exchange diagrams (dashed),
one, two, and three gluon exchange diagrams (dot-dashed),
up to nine gluon exchange diagrams (dot-dot-dashed),
and a complete set of ladder diagrams (solid).}
\end{figure}

Not only is the deficiency of the impulse approximation clear, it is
also essential that the {\em full} ladder of gluon exchanges be
included, especially for kinematics near $\pi$-$\pi$ resonances.  This
is displayed in more detail in Fig.~\ref{fig:129ladder} where the
effects from one, two, three, nine, and ``infinitely many''
$s$-channel gluon exchanges are compared.  To reach effective
convergence requires up to about 20 gluon exchanges away from the
resonance region.  However, near intermediate bound state poles
($\rho$ and $\sigma$) convergence at the 1\% level requires several
hundred gluon exchanges.

\begin{figure}[t]
\includegraphics[width=8cm]{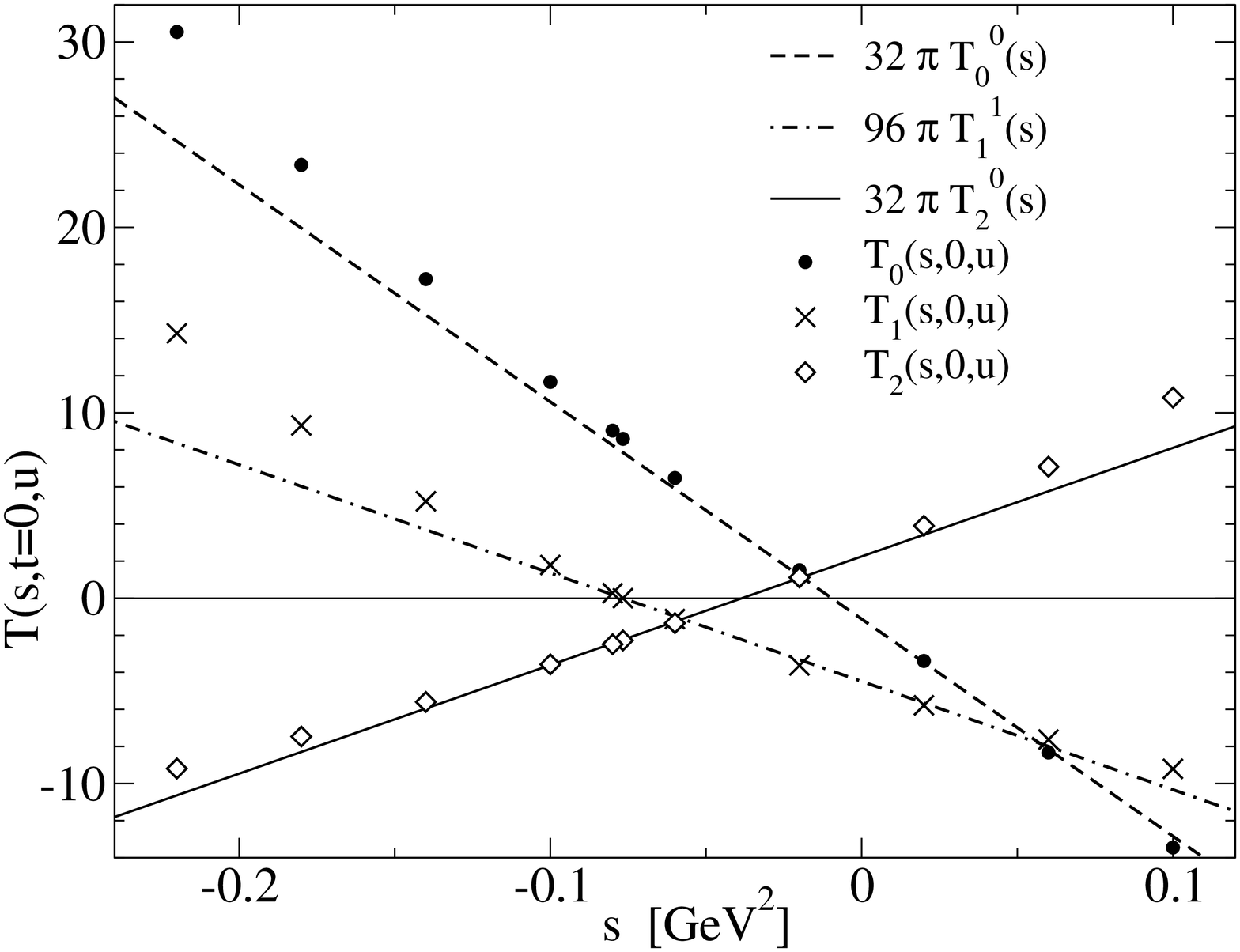}
\caption{\label{fig:tamps}
The lines correspond to leading-order chiral perturbation theory for
isospin zero (dashed), one (dot-dash), and two (solid) amplitudes near
threshold.  The symbols represent the DSBS amplitudes.}
\end{figure}

Combining the amplitudes $D$ and $E$ with
Eqs.~(\ref{eq:isospinT0})-(\ref{eq:isospinT2}) provides the different
isospin scattering amplitudes which are displayed in
Fig.~\ref{fig:tamps}.  For comparison we also plot the amplitudes from
leading-order chiral perturbation theory~\cite{Donoghue:1988xa} which
are given in partial wave form
\begin{eqnarray}
T_I = 32 \pi \sum_{L} (2 \; L + 1) T_I^L(s) P_L(\cos\theta) \ .
\end{eqnarray}
For low energies the $S$ and $P$ waves dominate and are given by
\begin{eqnarray}
T_0^0(s) &=& \frac{ -2\,s - m_\pi^2}{32 \pi\,f_\pi^2}
\label{eq:weinT0}	
\\
T_2^0(s) &=& \frac{ \,s + 2 m_\pi^2}{32 \pi\,f_\pi^2}
\label{eq:weinT1}	
\\
T_1^1(s) &=& \frac{ -\,s - 4 m_\pi^2}{96 \pi\,f_\pi^2 } \ .
\label{eq:weinT2}
\end{eqnarray}
Again recall that in the Euclidean metric $s = -4 m_\pi^2 = -0.076 \,
{\rm GeV}^2$ at threshold.

Note the excellent agreement between tree-level chiral perturbation
theory and our calculations at low energies.  The deviation at higher
energies represents contributions from $\pi$-$\pi$ resonances (scalar
and vector meson bound states).  These important physical effects are
automatically included in the present approach and are discussed
further in the next section.

Finally, we have calculated the dimensionless $S$ and $P$ wave
scattering lengths $a^L_I$ in all three isospin channels
\begin{eqnarray}
a_I^0 &=&  T_I^0(s = -4 m^2_{\pi})
 	\\
a_1^1 &=& \frac{ 4 m_\pi^2}{-s - 4 m_\pi^2} \;
	T_1^1(s \rightarrow -4 m^2_{\pi})
\end{eqnarray}
We find $a_0^0 = 0.17$, $a_1^1 = 0.036$, and $a_0^2 = -0.045$.  This
agrees well with Weinberg's theorem which is embodied in the
tree-level chiral perturbation results
\begin{eqnarray}
	a_0^0 &=& \frac{ 7 m_\pi^2}{32 \pi f_\pi^2} = 0.156
\\
	a_1^1 &=& \frac{ m_\pi^2}{24 \pi f_\pi^2} = 0.030
\\
	a_2^0 &=& \frac{ - 2 m_\pi^2}{32 \pi f_\pi^2} = -0.044 \; . 
\end{eqnarray}
A recent analysis~\cite{Colangelo:2001df} of the experimental
data~\cite{Pislak:2001bf} utilizes both two-loop chiral perturbation
theory and a phenomenological description involving the Roy
equations~\cite{Roy:1971tc} and obtains $a^0_0 = 0.220\pm 0.005$,
$a^1_1 = 0.0379\pm 0.0005$, and $a_2^0 = 0.0444 \pm 0.0010$.  Chiral
perturbation theory is able to provide more accurate scattering
lengths because they include higher order contributions from pion
loops which is especially important for the isospin zero channel:
one-loop chiral perturbation theory~\cite{Gasser:1984yg} gives $a^0_0
= 0.200$, compared to $a^0_0 = 0.156$ at leading order.  In our
present quark-based calculation pion loops are not included.  However,
they are clearly necessary for an accurate description of the data, in
particular in the isospin-zero channel.

\subsection{Comparison with meson-exchange models}
\begin{figure}[t]
\includegraphics[width=8cm]{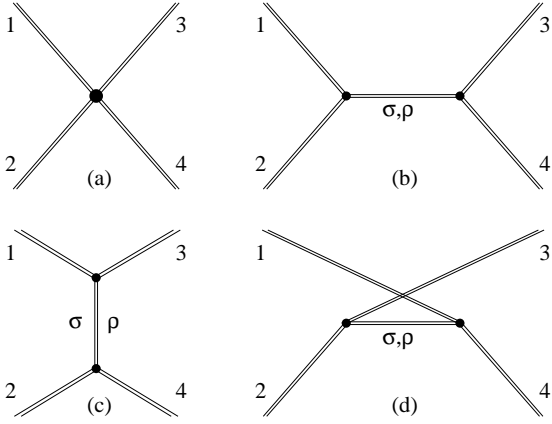}
\caption{\label{fig:QHD}
Leading meson-exchange diagrams for $\pi$-$\pi$ scattering: 
(a) contact term, (b) $s$-channel, (c) $t$-channel, and (d)
$u$-channel $\sigma$ and $\rho$ exchange.}
\end{figure}

It is enlightening to compare the DSBS approach to the QHD formalism
which has an established phenomenological legacy.  For example, one
could formulate a meson-exchange model with both scalar and vector
mesons starting from the effective Lagrangian
\begin{eqnarray}
\cal{L} &=&   \cal{L}_{\pi} + 
	\cal{L}_{\sigma} + \cal{L}_{\rho} + \cal{L}_{\text{int}} 
\\
\cal{L}_{\text{int}} &=& 
	g_{4\pi}\left(\vec\phi_\pi\cdot\vec\phi_\pi\right)^2 + 
	g_{\sigma \pi\pi} \phi_\sigma \vec\phi_\pi\cdot \vec\phi_\pi 
\nonumber \\ && {} +
	g_{\rho \pi \pi} \vec\phi^{\mu}_\rho \cdot \vec\phi_\pi 
		\times \nabla_\mu \vec\phi_\pi \;.
\end{eqnarray}
It entails meson masses and coupling constants which require
phenomenological determination; it also requires a certain amount of
fine-tuning in order to satisfy chiral constraints.  The leading
Feynman diagrams for $\pi$-$\pi$ scattering in such an approach are
displayed in Fig.~\ref{fig:QHD}.

In a meson-exchange model using both scalar and vector mesons, the
amplitude $D(s,t;u)$ corresponds to a contact term plus meson exchange
contributions in the $s$ and $t$ channels; the amplitude $E(s;t,u)$
corresponds to the same contact term plus meson exchange contributions
in the $t$ and $u$ channels.  In order to calculate these meson
exchange contributions, let us first specify the meson coupling
constants $g_{\sigma\pi\pi}$ and $g_{\rho\pi\pi}$, which in impulse
approximation are given by
\begin{eqnarray}
\lefteqn{ g_{\sigma\pi\pi} = }
\nonumber \\ && {}
	4\;N_c \int_q^\Lambda\! {\rm Tr}\Big[ 
	S(k+P_1) \Gamma_\pi(k+P_1,k) S(k) \Gamma_\pi(k,k-P_2) 
\nonumber \\ && {} \times 
	S(k-P_2) \Gamma_\sigma(k-P_2,k+P_1) \Big]
\\
\lefteqn{ g_{\rho\pi\pi} (P_1 - P_2)^\mu = }
\nonumber \\ && {}
	4\;N_c \int_q^\Lambda\! {\rm Tr}\Big[ 
	S(k+P_1) \Gamma_\pi(k+P_1,k) S(k) \Gamma_\pi(k,k-P_2) 
\nonumber \\ && {} \times 
	S(k-P_2) \Gamma^\mu_\rho(k-P_2,k+P_1) \Big]
\end{eqnarray}
for on-shell scalar and vector mesons, $(P_1 + P_2)^2 = -m_\sigma^2$,
$-m_\rho^2$ respectively.  

Within the present model, we have $m_\sigma^2 = 0.449~{\rm GeV}^2$,
$g_{\sigma\pi\pi} = 2.06~{\rm GeV}$, $m_\rho^2 = 0.549~{\rm GeV}^2$,
and $g_{\rho\pi\pi} = 5.14$~\cite{Jarecke:2002xd}.  The corresponding
width for the decay $\rho^0\to \pi^+\pi^-$ is
\begin{eqnarray}
 \Gamma_{\rho\pi\pi} &=& \frac{g^2_{\rho\pi\pi}}{48\,\pi}\; 
	\frac{(m_\rho^2 - 4 m_\pi^2)^{\frac{3}{2}}}{m_\rho^2}
\\
	&=& 104~{\rm MeV}
\end{eqnarray}
which is in reasonable agreement with the experimental $\rho$ width
$151~{\rm MeV}$; the difference could very well be explained by
pion loops which are not included in the present approach.
The decay width for the $\sigma\to \pi\pi$ is
\begin{eqnarray}
 \Gamma_{\sigma\pi\pi} &=& \frac{3}{2} \;
	\frac{g^2_{\sigma\pi\pi}}{16\,\pi} \;
	\frac{\sqrt{m_\sigma^2 - 4 m_\pi^2}}{m_\sigma^2}
\\
	&=& 172~{\rm MeV}
\end{eqnarray}
which is also quite reasonable for a broad $\sigma$ resonance (the
factor $\frac{3}{2}$ comes from summing over the charged and neutral
pions).  However, it is known that the scalar BSE receives significant
corrections beyond ladder truncation~\cite{Bender:1996bb}, which could
change this calculated decay width.  In addition, pion loops should be
incorporated self-consistently in the BSE approach for a more
realistic calculation of the $\sigma$ and $\rho$ widths.

Next, consider for example the contribution of $s$-channel ladder
diagrams to $D(s,t;u)$: scalar mesons contribute
\begin{eqnarray}
	\frac{g^2_{\sigma\pi\pi} / 4}{s + m_\sigma^2}
\end{eqnarray}
whereas vector mesons contribute
\begin{eqnarray}
	\frac{g^2_{\rho\pi\pi} \, (u-t) / 4}{s + m_\rho^2}
\end{eqnarray}
near the bound state poles, and similarly for the $t$- and $u$-channel
ladder diagrams.  The momentum-dependent factor in the numerator of
the vector meson exchange contributions comes from 
$(P_1 - P_2)\cdot(P_3 - P_4) = u-t$.

A detailed, consistent evaluation of these diagrams generates the
effective meson-exchange scattering amplitudes at tree level
\begin{eqnarray}
 T_0(s,t,u) &=& 5\,C_{4\pi} 
\nonumber \\ && {}
	+ \frac{1}{2} \, g_{\sigma\pi\pi}^2
	\left(\frac{3}{s + m_\sigma^2} 
		+ \frac{1}{t + m_\sigma^2} 
		+ \frac{1}{u + m_\sigma^2} \right)
\nonumber \\ && {}
	+ g_{\rho\pi\pi}^2
	\left(\frac{u - s}{t + m_\rho^2} 
		+ \frac{t - s}{u + m_\rho^2} \right)
\\
 T_1(s,t,u) &=& 
	\frac{1}{2}\,g_{\rho\pi\pi}^2
	\left(\frac{2(u - t)}{s + m_\rho^2} 
		+ \frac{u - s}{t + m_\rho^2} 
		- \frac{t - s}{u + m_\rho^2} \right)
\nonumber \\ && {}
	+ \frac{1}{2}\,g_{\sigma\pi\pi}^2
	\left(\frac{1}{t + m_\sigma^2} 
		- \frac{1}{u + m_\sigma^2} \right)	
\\
 T_2(s,t,u) &=& 2\,C_{4\pi}  
	+ \frac{1}{2}\,g_{\sigma\pi\pi}^2
	\left(\frac{1}{t + m_\sigma^2} 
		+ \frac{1}{u + m_\sigma^2} \right)
\nonumber \\ && {}
	- \frac{1}{2}\,g_{\rho\pi\pi}^2
	\left(\frac{u - s}{t + m_\rho^2} 
		+ \frac{t - s}{u + m_\rho^2} \right) \,.
\end{eqnarray}

We can now evaluate the accuracy of the meson exchange model by
comparison with the full DSBS amplitude.  A consistent assessment
requires that the meson parameters (i.e. the coupling constants and
meson masses) must be provided by the DSBS model.  The only free
parameter in the meson exchange model is $C_{4\pi}$, which can be
fitted such that the meson exchange model agrees with the microscopic
calculation away from the resonances.  A value of $C_{4\pi} \approx
-4.8$ provides reasonable agreement for both the isospin zero and two
results over a large kinematic region.  To reproduce Weinberg's
results for the scattering lengths $a_0^0$ and $a_2^0$ requires a
fine-tuning between the meson masses and coupling constants in the
meson exchange model.  The meson exchange model with calculated
$m_\rho$, $m_\sigma$, $g_{\rho\pi\pi}$, and $g_{\sigma\pi\pi}$ from
the DSBS model cannot reproduce the Weinberg limit: $a_0^0 = 0.156$
requires $C_{4\pi} = -5.2$, whereas $a_0^0 = -0.044$ requires
$C_{4\pi} = -4.0$.  In contrast, the DSBS approach does reproduce the
Weinberg limits correctly, while at the same time properly generating
non-analytic effects from intermediate $q\bar{q}$ bound states such as
$\rho$ and $\sigma$ mesons.

\begin{figure}[htb]
\includegraphics[width=8.5cm]{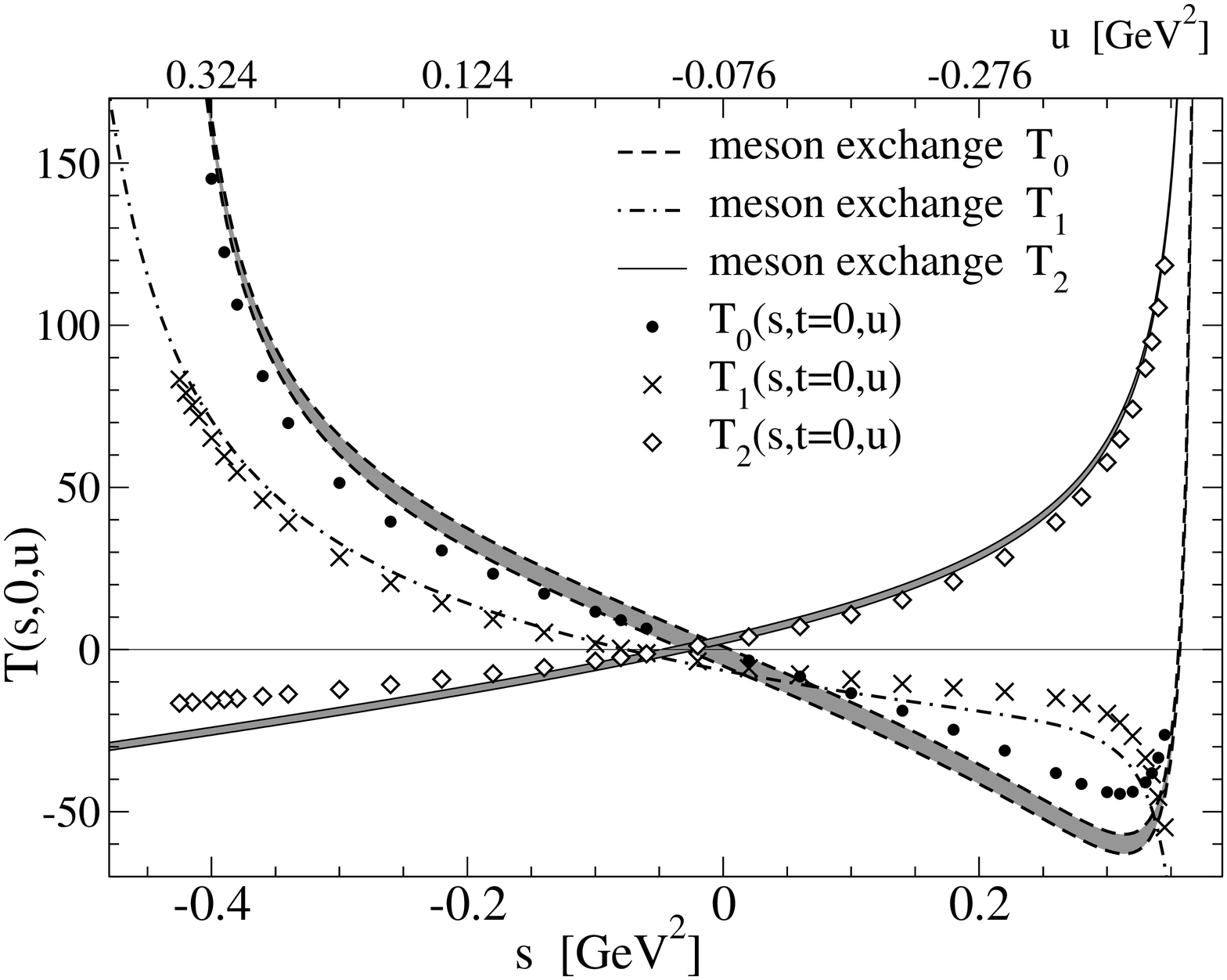}
\label{compare}
\caption{\label{fig:compare}
Isospin zero, one, and two $\pi$-$\pi$ amplitudes.  The symbols are
the DSBS calculations.  The lines are meson exchange model results,
using calculated meson masses and couplings from the DSBS approach.}
\end{figure}
Fig.~\ref{fig:compare} summarizes our comparative study for all three
isospin amplitudes at forward scattering.  The gray band indicates the
sensitivity on $C_{4\pi}$ for $-4.0 < C_{4\pi} < -5.2$.  Notice that
the $\sigma$ resonance naturally appears in the DSBS $S$ wave, isospin
zero amplitude and agrees with the meson exchange amplitude quite
accurately near the $\sigma$ mass region.  Similarly the DSBS $P$
wave, isospin one amplitude is reasonably well reproduced by the meson
exchange result, although we cannot numerically quite reach the $\rho$
pole. 
In between threshold and resonant regions the two
approaches have some quantitative differences, however we conclude
that the tree-level meson exchange approach is reasonable for simple
predictions.

\section{Conclusion}
Summarizing, we have performed a consistent relativistic quark
formulation of $\pi$-$\pi$ scattering using the Dyson--Schwinger,
Bethe--Salpeter framework.  The DSBS approach provides contact with
QCD and naturally includes the important features of gauge and
Poincare invariance as well as both crossing and chiral symmetry.  We
have obtained numerical results which are in good agreement with the
observed isospin scattering lengths and reproduce Weinberg's theorem
at threshold.  Perhaps even more significant is the emergence of the
$\sigma$ and $\rho$ resonances at the proper energy which reaffirms
the QCD structure elements in the model.  The latter permits a
rigorous assessment of QHD meson exchange models which simply insert
resonances phenomenologically.  As we documented in detail, a simple
QHD treatment appears reasonable, especially near resonances.  Future
work will confront data at higher energies and also address other
flavored systems such as $\pi$-$K$ scattering.

\begin{acknowledgments}
This work was supported by the Department of Energy under Grants
DE-FG02-96ER40947 and DE-FG02-97ER41048. Calculations were performed
with resources provided by the National Energy Research Scientific
Computing Center and the North Carolina Supercomputer Center.
\end{acknowledgments}

\appendix*
%
\section{Isospin decomposition}
In terms of the $u,d$ flavor components, the
single pion states are
\begin{eqnarray}
  |\pi^+\rangle &=& |1,1\rangle = -|u\bar{d}\rangle	\\
  |\pi^0\rangle &=& |1,0\rangle =
	|(u\bar{u} - d\bar{d})/\sqrt{2}\rangle		\\
  |\pi^-\rangle &=& |1,-1\rangle = |d\bar{u}\rangle \ .
\end{eqnarray}

The two-pion states then have isospin decomposition
\begin{eqnarray}
  |\pi^\pm\pi^\pm\rangle &=& |2,\pm2\rangle 			\\
  |\pi^\pm\pi^0\rangle &=& \frac{1}{\sqrt{2}}
	\left(|2,\pm1\rangle \pm|2,\pm1\rangle \right)		\\
  |\pi^\pm\pi^\mp\rangle &=&
	\frac{1}{\sqrt{6}}|2,0\rangle \pm
	\frac{1}{\sqrt{2}}|1,0\rangle +
	\frac{1}{\sqrt{3}}|0,0\rangle 				\\
  |\pi^0\pi^0\rangle &=&
	\frac{2}{\sqrt{3}}|2,0\rangle -
	\frac{1}{\sqrt{3}}|0,0\rangle \ .
\end{eqnarray}
Using the latter equations, the physical $\pi$-$\pi$ scattering
amplitudes can be related to the three isospin amplitudes by
\begin{eqnarray}
\langle\pi^\pm\pi^\pm|S|\pi^\pm\pi^\pm\rangle &=& T_2		\\
\langle\pi^\pm\pi^\mp|S|\pi^\pm\pi^\mp\rangle &=&
	\frac16 T_2 + \frac12 T_1 + \frac13 T_0 	 	\\
\langle\pi^\pm\pi^0|S|\pi^\pm\pi^0\rangle &=&
	\frac12 T_2 + \frac12 T_1				\\
\langle\pi^\pm\pi^\mp|S|\pi^0\pi^0\rangle &=&
	\frac13 T_2 - \frac13 T_0 	 			\\
\langle\pi^0\pi^0|S|\pi^0\pi^0\rangle &=&
	\frac23 T_2 + \frac13 T_0 \ .
\end{eqnarray}

Not all six diagrams in Fig.~\ref{fig:conf} contribute to each
$\pi$-$\pi$ scattering amplitude.  In terms of the physical
pion scattering amplitudes, we have
\begin{eqnarray}
\langle\pi^\pm\pi^\pm|S|\pi^\pm\pi^\pm\rangle &=&
	E(s,t,u) + E(s,u,t) 					\\
\langle\pi^\pm\pi^\mp|S|\pi^\pm\pi^\mp\rangle &=&
	2\;D(s,t,u)						\\
\langle\pi^\pm\pi^0|S|\pi^\pm\pi^0\rangle &=&
	D(s,t,u) - D(s,u,t)
\nonumber \\ && {}
	+ \frac{1}{2}\left(E(s,t,u) + E(s,u,t)\right) 		
\nonumber\\
\\
\langle\pi^\pm\pi^\mp|S|\pi^0\pi^0\rangle &=&
	- D(s,t,u) - D(s,u,t)
\nonumber \\ && {}
	+ \frac{1}{2}\left(E(s,t,u) + E(s,u,t)\right) 		
\nonumber\\
\\
\langle\pi^0\pi^0|S|\pi^0\pi^0\rangle &=&
	D(s,t,u) + D(s,u,t)
\nonumber \\ && {}
	+ \frac{1}{2}\left(E(s,t,u) + E(s,u,t)\right) \ .
\nonumber\\
\end{eqnarray}
Thus we find for the isospin amplitudes
\begin{eqnarray}
T_0(s,t,u) &=& 3 \left( D(s,t,u) + D(s,u,t) \right)\
\nonumber \\ && {}
		- \frac{1}{2}\left(E(s,t,u) + E(s,u,t)\right) 	\\
T_1(s,t,u) &=& 2 \left( D(s,t,u) - D(s,u,t) \right)		\\
T_2(s,t,u) &=& E(s,t,u) + E(s,u,t) \ .
\end{eqnarray}
The quark exchange diagram, $E(s;t,u)$, is symmetric in the last two
arguments, $t$ and $u$, whereas the quark annihilation diagram,
$D(s,t;u)$, is symmetric in the first two arguments, $s$ and $t$.
Furthermore, the quark exchange and quark annihilation diagram are
related to each other via
\begin{eqnarray}
  E(s;t,u) &=& D(u,t;s)  \ .
\end{eqnarray}
Thus, there is actually only one independent amplitude, say
$D(s,t;u)$, where the symbol $;$ denotes the amplitude is symmetric in
the first two arguments.  In terms of this single amplitude, we have
\begin{eqnarray}
	T_0(s,t,u) &=& 3 \; D(s,t;u) + 3 \; D(s,u;t)
\nonumber \\ && {}
			- D(u,t;s)			
\label{eq:T0}				\\
	T_1(s,t,u) &=& 2 \; D(s,t;u) - 2 \; D(s,u;t) 	
\label{eq:T1}				\\
	T_2(s,t,u) &=& 2 \; D(u,t;s) \;.
\label{eq:T2}
\end{eqnarray}
Finally, we recall the more conventional Weinberg amplitudes $A$, $B$,
and $C$ defined by the standard scattering amplitude relation
\begin{eqnarray}
\lefteqn{\langle\pi^\gamma\pi^\delta|S|\pi^\alpha\pi^\beta\rangle =
 A(s,t,u)\delta_{\alpha \beta}\delta_{\gamma \delta} } 
\nonumber \\ && \hbox{} +
 B(s,t,u)\delta_{\alpha \gamma}\delta_{\beta \delta} +
 C(s,t,u)\delta_{\alpha \delta}\delta_{\beta \gamma} \ .
\end{eqnarray}
Here the indices $\alpha, \beta, \gamma $ and $\delta$ refer to the
cartesian isospin projections ($x,y,z$) for the pion, which are
related to the physical charge states by
\begin{eqnarray}
|\pi^\pm\rangle &=& 
	\mp \Big(|\pi^x\rangle  \pm i |\pi^y\rangle\Big)/\sqrt{2} \\
|\pi^0\rangle &=& |\pi^z\rangle  \ .
\end{eqnarray}
An elementary isospin calculation immediately yields
\begin{eqnarray}
T_0(s,t,u) &=& 3 \; A(s,t,u) +  \; B(s,t,u)
\nonumber \\ && {}
		+ C(s,t,u)			\\
\label{weinbergT0}
T_1(s,t,u) &=&  \; B(s,t,u) -  \; C(s,t,u) 	\\
\label{weinbergT1}
T_2(s,t,u) &=&  \; B(s,t,u) +  \; C(s,t,u) \ .
\label{weinbergT2}
\end{eqnarray}
Then comparing with Eqs.~(\ref{eq:T0})-(\ref{eq:T2}) we can
relate the three Weinberg amplitudes to the DSBS amplitude $D$
\begin{eqnarray}
	A(s,t,u) &=& D(s,t;u) + D(s,u;t) - D(t,u;s)
\label{weinbergA}						\\
	B(s,t,u) &=& D(s,t;u) - D(s,u;t) + D(t,u;s) 		\\
	C(s,t,u) &=& -D(s,t;u) + D(s,u;t) + D(t,u;s) \ .	
\nonumber\\
\end{eqnarray}
Since the Weinberg amplitudes satisfy
\begin{eqnarray}
	B(s,t,u) &=& A(t,s,u)				\\
	C(s,t,u) &=& A(u,t,s)
\end{eqnarray}
we can also use Eq.~(\ref{weinbergA})  to again derive $B$ and $C$
\begin{eqnarray}
	B(s,t,u) = D(s,t;u) + D(t,u;s) - D(s,u;t) 	\\
	C(s,t,u) = D(u,t;s) + D(s,u;t) - D(s,t;u)
\end{eqnarray}
as an independent check.

\bibliography{refsPM,refsPCT,refsCDR,refs,refspipi}

\end{document}